# ACCELERATING SYSTEM VERILOG UVM BASED VIP TO IMPROVE METHODOLOGY FOR VERIFICATION OF IMAGE SIGNAL PROCESSING DESIGNS USING HW EMULATOR

Abhishek Jain[1], Piyush Kumar Gupta[2], Dr. Hima Gupta[3] and Sachish Dhar [4]

[1]Imaging Group, STMicroelectronics, Greater Noida, India
[1,3]JBS, Jaypee Institute of Information Technology (JIIT), Noida, India
[2,4]SDS Group, STMicroelectronics, Greater Noida, India

## ABSTRACT

*In this paper we present the development of Acceleratable UVCs from standard UVCs in System Verilog and their usage in UVM based Verification Environment of Image Signal Processing designs to increase run time performance. This paper covers development of Acceleratable UVCs from standard UVCs for internal control and data buses of ST imaging group by partitioning of transaction-level components and cycle-accurate signal-level components between the software simulator and hardware accelerator respectively. Standard Co-Emulation API: Modeling Interface (SCE-MI) compliant, transaction-level communications link between test benches running on a host system and Emulation machine is established. Accelerated Verification IPs are used at UVM based Verification Environment of Image Signal Processing designs both with simulator and emulator as UVM acceleration is an extension of the standard simulation-only UVM and is fully backward compatible with it. Acceleratable UVCs significantly reduces development schedule risks while leveraging transaction models used during simulation.*

*In this paper, we discuss our experiences on UVM based methodology adoption on TestBench-Xpress (TBX) based technology step by step. We are also doing comparison between the run time performance results from earlier simulator-only environment and the new, hardware-accelerated environment. Although this paper focuses on Acceleratable UVC's development and their usage for image signal processing designs. Same concept can be extended for non-image signal processing designs.*

## KEYWORDS

*SystemVerilog, Universal Verification Methodology (UVM), TestBench-Xpress (TBX), Universal Verification Component (UVC), Standard Co-Emulation API: Modelling Interface (SCE-MI), Acceleratable UVC, Emulator, XRTL Tasks/Functions (xtf), Transactor interface (tif), Verification IP (VIP).*

## 1. INTRODUCTION

Universal Verification Methodology (UVM) is a rich and capable class library that has evolved over several years from much experience with real verification projects large and small, and SystemVerilog itself is a large and complex language. As a result, although UVM offers a lot of powerful features for verification experts, it can present a daunting challenge to Verilog and VHDL designers who want to start benefitting from test bench reuse [5].

TestBench-Xpress (TBX) technology delivers the same functionality achievable in simulation with advanced and simple debug capabilities, but at 10-1000x of times faster performance.

Additionally, it greatly increases verification productivity by using the same testbench for simulation and acceleration [16].

Usually in case of co-emulation with TBX technology where non-synthesizable HVL part mapped on Host Machines communicates with HDL part which is mapped on Emulators through SCEMI [14] or TBA standards wins race based on High performance efficient approach.

This paper describes development of Acceleratable UVCs from standard UVCs in System Verilog and their usage in UVM based Verification Environment of Image Signal Processing designs to increase run time performance. The Image signal processing algorithms are developed and evaluated using Python models before RTL implementation. Once the algorithm is finalized, Python models are used as a golden reference model for the IP development. To maximize re-use of design effort, the common bus protocols are defined for internal register and data transfers. A combination of such configurable image signal processing IP modules are integrated together to satisfy a wide range of complex video processing SoCs [1], [2].

Verification Environment of Image Signal Processing IP and Sub-System is described in detail in Section 4.

## 2. EMULATION APPROACH

Hardware Emulation has been matured enough in Industry used as an integral part of life-cycle for any SoC and IP verification. As design is becoming more and more complex and moving towards Multi-million to Billion gates size, emulation provides accelerated simulation environment to help verification engineer finding bugs quite before silicon. In spite of being slow from traditional FPGA prototyping, Emulation is getting increasingly famous for pre-silicon validation where software engineers are able to develop software applications, boot Linux on SoC etc. much before Silicon.

In-Circuit-Emulation mode is a traditional way where testbench and DUT both are synthesizable and mapped on Hardware Emulator box to have faster performance. The same platform can be used by Software engineers for pre-silicon validation. Software debug connections to emulation have traditionally been handled using hardware-based, JTAG probe connections. Because JTAG uses a serial data connection, performance is limited on the emulator.

In Cycle Accurate Co-emulation, the testbench is written and executed in HVL for greater testbench performance. Signals are synchronized at clock boundaries. Clocks advance under control of the HVL testbench. This approach makes complete system slower as there will always be interaction with Hardware and Software at each clock.

In Transaction-Level Co-Emulation, the testbench is written in SystemC, C++ or SystemVerilog. Packets of data (transactions) are exchanged between the testbench and the DUT. This reduces the communication time between the host machine and emulator as data transfers are performed in transaction level instead of signal level first approach. To do this, transactor should be described in a synthesizable way to mapped on hardware emulator with DUT. Moreover, the transactor design depends on both emulator system protocol and DUT protocol. Therefore, transactor description would not only be time-consuming but also error-prone task [15].

### 2.1 SCE-MI INTERFACE

The Standard Co-Emulation Modelling Interface (SCE-MI) was first introduced at that time as a way to standardize the communication between the hardware portion running in the emulator and the software portion running on the Host Machine [14].

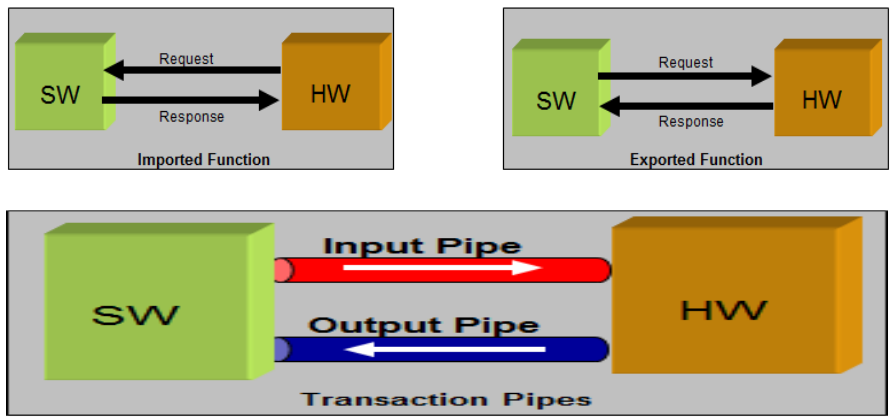

Figure 1: SCEMI-2 Infrastructure

## 2.2 TRANSACTION-BASED ACCELARATION

TBX establishes a SCE-MI compliant, transaction-level communications link between testbenches running on a host system and SoC mapped on Veloce hardware emulation box.

Transaction-level verification is a verification methodology both in simulation and emulation. In emulation it is further leveraged due to the superior performance that it yields. Transactors are an important component in transaction-level verification, and serve as the bridge between a test environment written in a Hardware Verification Language (HVL) and the DesignUnderTest (DUT) inside the Veloce emulator. The Transactor is responsible for converting the high-level HVL commands into low-level DUT pin wiggles (HDL), and handling the communication between the two domains (HVL and HDL) (Figure 2) [16], [17].

A protocol transactor implements a protocol (AMBA, USB, ST Internal Protocol and so on) which drives the DUT interface in a protocol-compliant way, and captures DUT responses into high level protocol transactions.

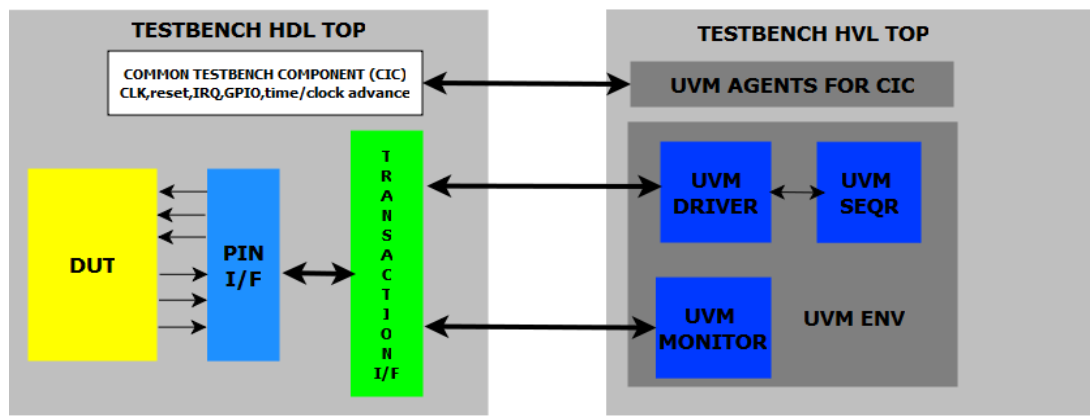

Figure 2: Transactor Bridging from HVL to HDL

Due to the high-level interface at the HVL side, the verification environment is free from modelling low-level protocol details, thus ensuring easy and more comprehensive test development. This transaction-level verification environment can now run at full emulation performance using Testbench-Xpress (TBX) and Veloce, without sacrificing much of the functional coverage of the protocol [16], [17].

We have developed and used ST internal control and data bus Accelerated VIPs in our IP and Subsystem level Verification Environment and also using standard Acceleratable UVC's in SoC Level Verification.

## 3. TBX TECHNOLOGY WITH UVM METHODOLOGY

A transaction can be defined as a transfer of data from one component to another that may or may not consume time [12]. In any procedural language like C, SystemC or SystemVerilog, a transaction is equivalent to a function call. TBX facilitates this through its support of remote procedure invocation, whereby, tasks or function calls defined on one domain could be called from the other.

For running on TBX, the environment must be partitioned into synthesizable XRTL compliant HDL files and the HVL files containing the high-level test bench components and compiled separately. So it will not be always needed to synthesize the HDL side which is time consuming.

Any transaction passed from HVL and HDL layers, via an xtf, must be packed into an equivalent static packed data structure that could be synthesized by TBX. Similarly, the XRTL will send a packed data structure that can be unpacked by the HVL to create transaction objects.

For such a scenario, it is best to divide the actual HVL transactor into a synthesizable XRTL transactor interface (tif) and a non-synthesizable proxy class. The tif can have a handle to the proxy class. The proxy class can contain a handle to the actual tif. The tif can call functions defined in the proxy, and the proxy can call functions or tasks defined in the tif [16].

In below sections, we will describe the verification environment of Image signal Processing IP and Subsystem and steps followed to convert standard UVC's into Acceleratable UVC's.

## 4. VERIFICATION ENVIRONMENT OF IMAGE SIGNAL PROCESSING IP AND SUB-SYSTEM

In an image signal processing IP as shown in figure 3, there are A input video data interfaces, C output video data interfaces, B memory interfaces, D output Interrupts and E register interfaces, where A, B, C, D and E values can be from 0 to any arbitrary number [1].

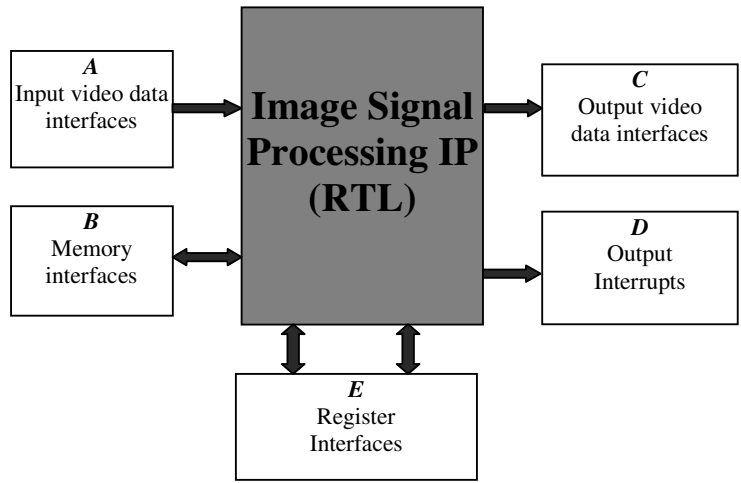

Figure 3: Image Signal Processing IP Block Diagram

At subsystem level, all the R IP's in Image signal processor pipe are connected serially. Generally output data interface of one IP is connected to the input data interface of another IP as shown in figure 4.

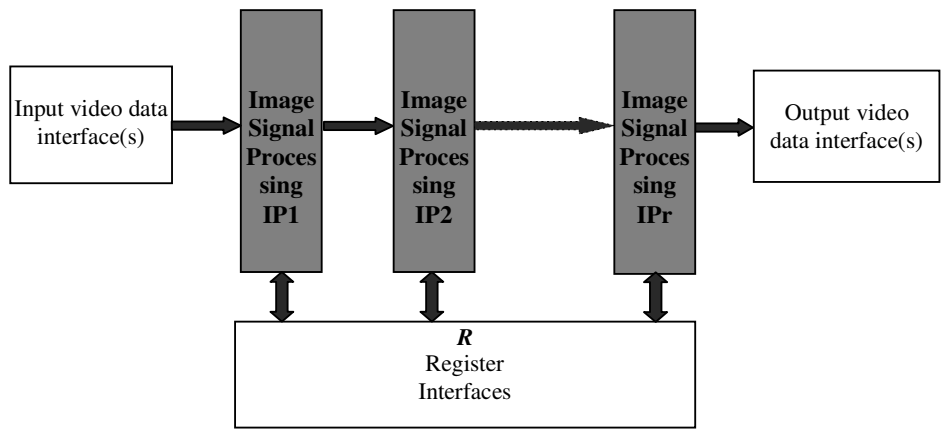

Figure 4: Image Signal Processing Subsystem Block Diagram

For verifying interfaces of an image signal processing IP, dedicated UVCs are used. In case of register interface(s), register interface UVC and UVM_REG register model are used. Similarly for video data interface(s), video data interface UVC is used.

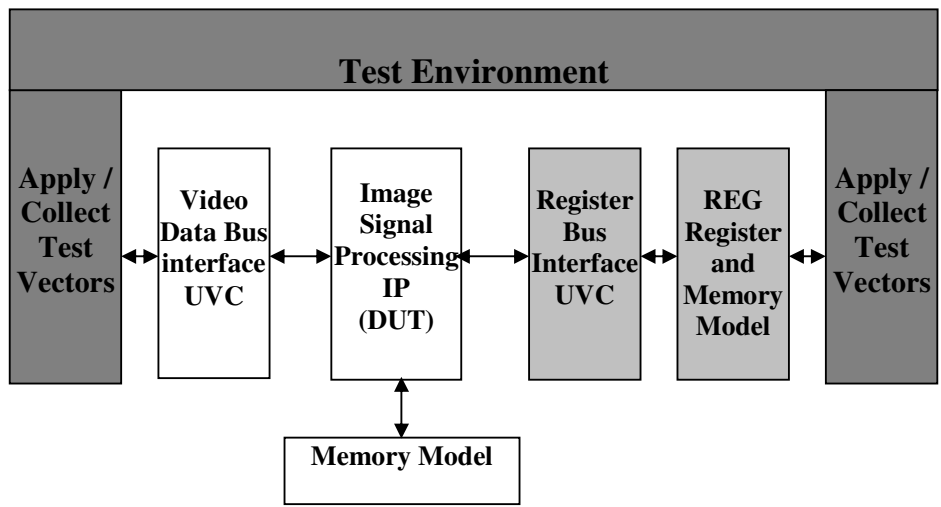

Figure. 5 Basic blocks of System Verilog UVM based IP Level Verification Environment

Note that there can be multiple instances of these UVC's in a verification environment. Each agent is configured separately and any combination of agent configurations can coexist in the same environment. Therefore in above case, *E* instances of register interface UVC agents and M (M = max (A, C)) instances of video data interface UVC agents are used to interface with a DUT. Figure 5 illustrates the basic blocks of System Verilog UVM based IP Level Verification Environment [1].

Similar to IP Level Verification Environment, for verifying subsystem of image signal processor also, we use internal video data interface UVC for video data interface and register interface UVC and UVM_REG register model for register interface(s) as shown in figure 6.

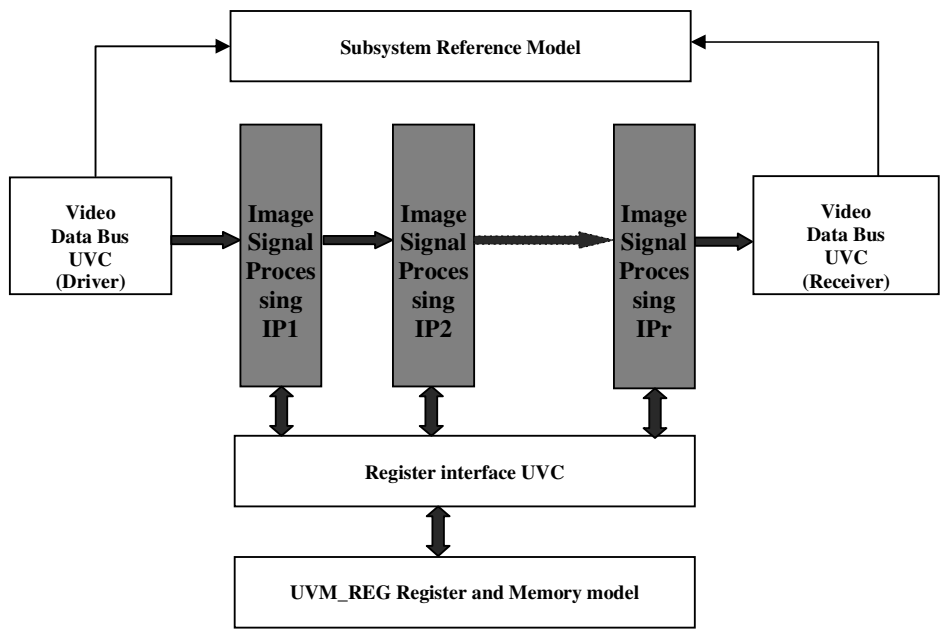

Figure. 6 System Verilog UVM based Subsystem Level Verification Environment

International Journal of VLSI design & Communication Systems (VLSICS) Vol.4, No.6, December 2013

## 4.1 Development of Acceleratable UVCs from standard UVCs

A Simulation based Verification IP (VIP) is a SystemVerilog interface driving a DUT interface (pin connections or SystemVerilog interface) on one side and connected to a test bench environment on the other side (like SV, OVM, or UVM), through a transaction-based set of APIs [3], [13].

Figure 7 shows Register Bus UVC's environment.
- Constrained random generation of protocol stimulus and driven through the API layer into the model. The model converts this high-level transaction into pin wiggles on the DUT interface.
- The model also captures responses from the interface (bus) and creates a high-level transaction the monitor recognizes on the test bench side. The monitor sends it to the various analysis ports where coverage and scoreboard modules are connected.

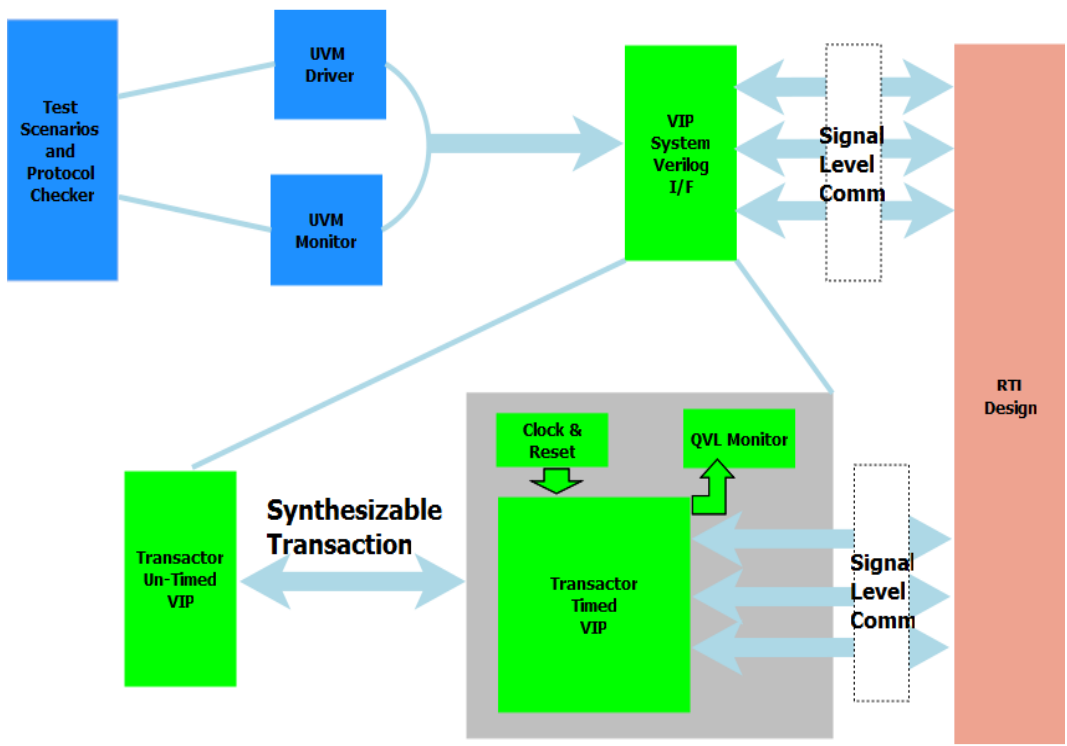

Figure. 7 Comparison of Veloce Transactor to the Simulation Based VIP

Figure -7 gives a look for the Veloce Transactor comparison to the Simulation Based Control Bus VIP. The verification environment in Veloce is in two domains: the XRTL (timed) portion of the transactor in Veloce, and the HVL (untimed) portion in the workstation (software). Models described using high-level language (HVL) constructs are executed by the simulator and the models described using hardware description language (HDL) constructs are executed by the hardware accelerator. Clocks and Reset are part of timed component and can be generated using TBX clkgen pragma which allow tool to synthesize this behavioral code and make it reside on Emulators [17], [18].

A UVM agent generally contains sequencer, driver, and monitor.

A sequence item is a transaction object from the sequencer that stimulates the driver [11]. In order to transfer a data item from the proxy in the HVL portion to the BFM in the HDL portion, the data members need to map into a packed struct Packet_t. Figure-8 shows the modelling of a sequence item Packet and a corresponding SystemVerilog packed struct Packet_t which represents synthesizable transaction of Packet.

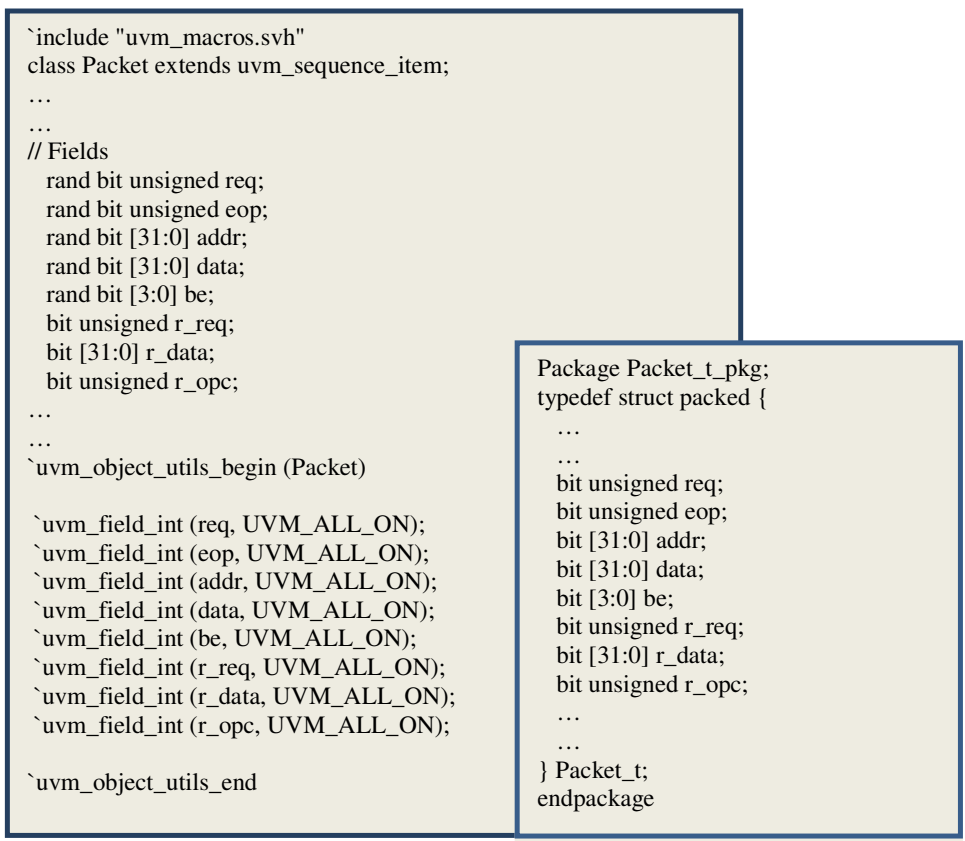

Figure 8: two representations of transaction

The class-based control bus UVC's driver receives a sequence item, converts it to a SystemVerilog struct, and passes the transaction referred by a virtual interface. For conversion between the two representations (mentioned above), we need to declare function "from_class_to_struct" in driver class. In this model, the bus functional models (BFM) which are XRTL tasks/functions to drive DUT pins are implemented in a synthesizable SV Transactor interface. During the connect phase, the virtual interface of the UVM driver connects to a virtual interface BFM (in Figure-9) which, at the end of the elaboration step, connects to the actual transaction interface instance (driver_bfm_if).

```
import Packet_t_pkg::*;
class driver_proxy extends uvm_driver # (Packet);
Packet_t req_s;
//virtual driver_interface
virtual driver_bfm_if BFM;
........................

//build phase to get virtual interface
virtual function void build_phase (uvm_phase phase);
super.build_phase(phase);
uvm_config_db #(virtual driver_bfm_if)::get(this,"","driver_bfm_if",BFM);
if(BFM == null)
begin
`uvm_fatal("DRIVER_INTERFACE CONFIG ERROR", "driver_bfm_inf is not set in driver proxy class");
end
endfunction

function Packet_t from_class_to_struct(Packet Packet_c);
Packet_t Packet_t_s;
  Packet_t_s.req      =    Packet_c.req;
  Packet_t_s.eop      =    Packet_c.eop;
  Packet_t_s.addr     =    Packet_c.addr;
  Packet_t_s.data     =    Packet_c.data;
  Packet_t_s.be       =    Packet_c.be;
  Packet_t_s.r_req    =    Packet_c.r_req;
  Packet_t_s.r_data   =    Packet_c.r_data;
  Packet_t_s.r_opc    =    Packet_c.r_opc;
return Packet_t_s;
endfunction

//execute the run phase
virtual task run_phase(uvm_phase phase);
//BFM.wait_for_reset();
forever
begin
  seq_item_port.get_next_item(req);
  req_s = from_class_to_struct(req);
  BFM.drive(req_s);
  seq_item_port.item_done();
  ..............................

endclass
```

```
Interface driver_bfm_if ();
//pragma attribute driver_bfm_if partition_interface_xif
import Packet_t_pkg::*;

task drive (input stimulus_s my_packet );//pragma tbx xtf
 @ (posedge clk);
 begin
         req     = my_packet.req;
         ....................
 end
endtask
endinterface
```

Figure 9: Driver Proxy and Driver Interface

The synthesizable transaction interface (driver_bfm_if) contains functions and tasks to apply transaction packets to DUT pins. It contains tasks that a UVM driver uses to write the transaction item. Figure 9 shows the connection of an actual interface to a virtual interface and its connection to the driver.

Below is Monitor implementation (Figure 10) of Control Bus UVC, where transaction interface (monitor_bfm_if) contains task to apply DUT pins into transaction item. Interface task have a proxy function call to transfer synthesizable transaction to proxy side monitor where proxy side uses conversion from System Verilog struct to class sequence item type that can be further used for scoreboarding and other purposes [17], [18].

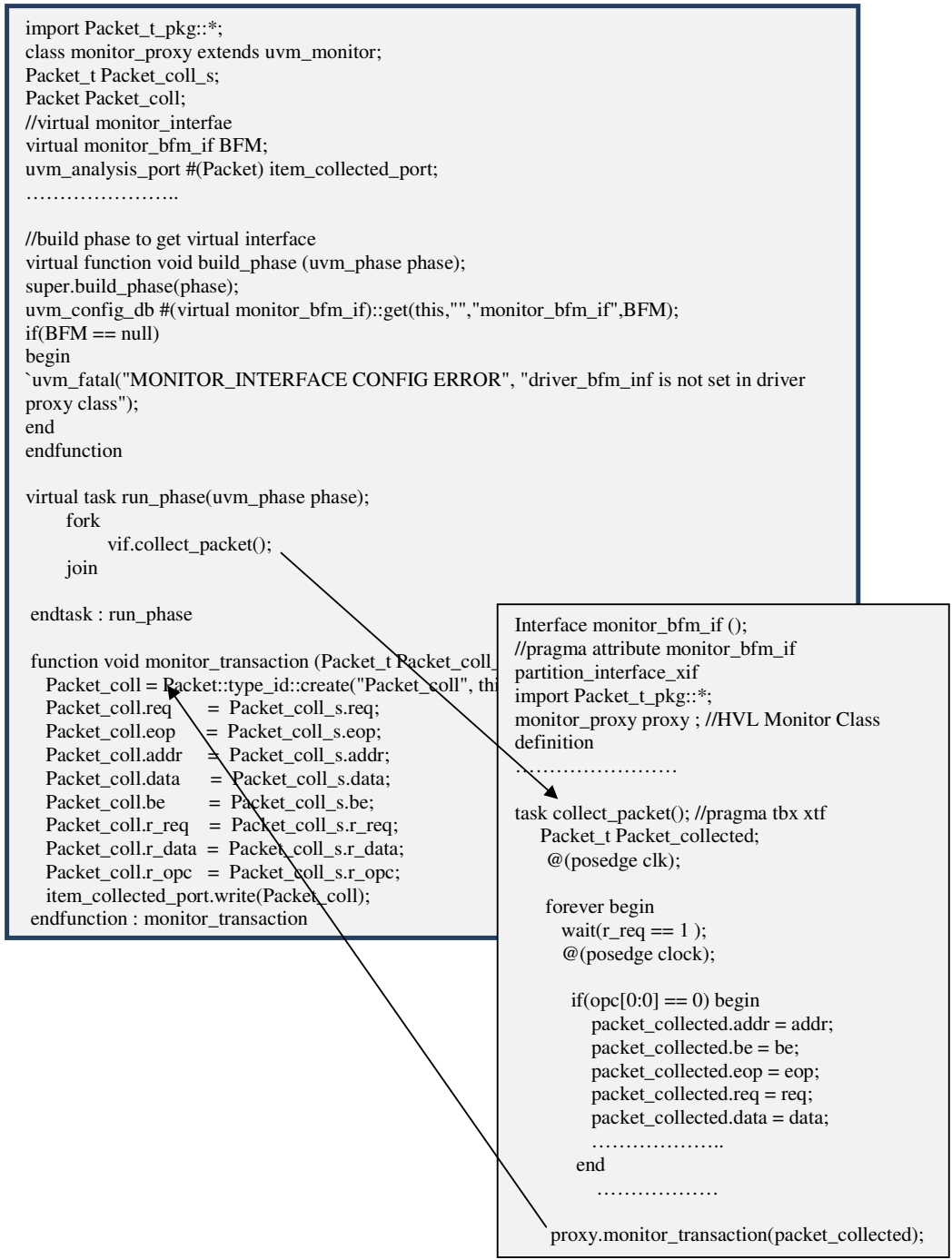

Figure 10: Monitor Proxy and Monitor Interface

As shown in figure 11, The Virtual Interface Binding can be done easily at HVL top for concrete interface instances. This complete model is native SystemVerilog and hence works in any SystemVerilog compliant simulator [6], [19].

```
module test;
import uvm_pkg::*;
`include "uvm_macros.svh"
`include "register_test.sv"

initial begin
 uvm_config_db#(virtual driver_bfm_if )::set(null, "uvm_test_top.env.i_agent.drv", "driver_bfm_if",
testbench_hdl_top.DRIVER_BFM);
uvm_config_db#(virtual monitor_bfm_if )::set(null, "uvm_test_top.env.i_agent.mon",
"monitor_bfm_if", testbench_hdl_top.MONITOR_BFM);
run_test("register_test");
end
endmodule
```

Figure 11: Virtual Interface Binding

UVM_REG register and memory model is used to write register/memory sequences that access hardware registers and memory areas and thus, it is used as generator in verification environment and is independent of the DUT interface. UVM_REG Register and memory model is described using high-level language (HVL) constructs and is executed by the simulator [10].

In the similar way, video data bus Acceleratable UVC from video data bus standard UVC is developed and used in Image Signal Processing designs.

### 4.2 Guidelines and Performance

To implement the unified testbench for simulation and acceleration, we followed the following coding guidelines:
- # Delays are not allowed in the testbench code.
- To achieve best performance, all code on the HVL testbench side must be untimed, and all timed code should be synthesized.
- There should not be any direct signal access from the HVL side. All communication must be transaction based.

## 5. RESULTS

Performance figure for Simulation vs. Emulation is described in below table.

Table 1: Performance comparison

| Design Size | Simulator time(Seconds) | Co-Emulation time(Seconds) | Gain in Co-emulation over simulation |
|---|---|---|---|
| ~5M gate | ~657 | ~20.44 | ~30X |
| ~9.5M gate | ~2044 | ~50.27 | ~40X |

## 6. CONCLUSIONS

In this paper, we presented the usage of fast growing UVM based unified co-emulation approach in image signal processing designs. Development of Unified Acceleratable UVCs from standard UVCs reduced development schedule risks while leveraging transaction models used during simulation. The key architecture and implementation specific decision for this acceleration VIP are made to maximise the reuse of same tests in simulation and acceleration platforms. This unified approach eliminated the penalty related to maintain different verification components for different platforms.

Additionally, The completeness of this setup – use of same accelerated VIP with simulator and emulator in UVM based Verification Environment gave us complete confidence that extension of the standard simulation-only UVM to include hardware acceleration will make verification of chips more productive. This approach take advantage of very fast emulator performance to handle longer and more regressive tests to cover more design areas and uncovering design bugs. This translates to regression tests that took hours to run in simulation are now taking few minutes to run on emulators.

In our Imaging designs environment, some of the results which we had listed in table 1 [Section 5] are example where we have taken different design setup to run with simulator as well as on emulator to estimate performance gain. While running a design of ~5Million gate size on these platforms, a specific testcase on simulator was taking nearly ~657 seconds compare to ~20 seconds on emulator which shows significant performance gain of ~30X. In another case, a different design of ~9.5 Million gate size have the performance gain of ~40X. This performance gain can be further improved by adopting more efficient combination of streaming and reactive transactions in future.

In emulation, code coverage is not completely supported so we mainly focused on functional coverage. Using Acceleratable VIP, we achieved approximately same functional coverage goal as with the standard simulation-only VIP.

## ACKNOWLEDGEMENTS

The authors would like to specially thank to their management Giuseppe Bonanno (CAD Manager, Imaging Division, STMicroelectronics) and Tran NGUYEN (Manager, SDS Emulation Team ) for their guidance and support. We would also like to thank management and team members of Imaging Division, STMicroelectronics; Faculty members and peer scholars of JBS, Jaypee Institute of Information Technology University and also Mentor Graphics team for their support and guidance.

## AUTHORS

**Abhishek Jain, Technical Manager, STMicroelectronics Pvt. Ltd.**
**Research Scholar, JBS, Jaypee Institute of Information Technology, Noida, India.**
**Email:** ajain_design@yahoo.co.in;
 abhishek-mmc.jain@st.com

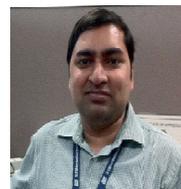

Abhishek Jain has more than 11 years of experience in Industry. He is driving key activities on Functional Verification Flow in Imaging Division of STMicroelectronics.
He has done PGDBA in Operations Management from Symbiosis, M.Tech in Computer Science from IETE and M.Sc. (Electronics) from University of Delhi. His main area of Interest is Project Management, Advanced Functional Verification Technologies and System Design and Verification especially UVM based Verification, Emulation/Acceleration and Virtual System Platform. Currently he is doing Research in Advanced Verification Methods for Efficient Verification Management in Semiconductor Sector. Abhishek Jain is a member of IETE (MIETE).

**Dr. Hima Gupta, Associate Professor, Jaypee Business School (A constituent of Jaypee Institute of Information Technology University), A – 10, Sector-62, Noida, 201 307 India.**
**Email:** hima_gupta2001@yahoo.com

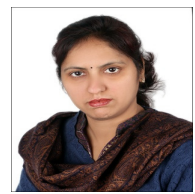

Dr. Hima has worked with LNJ Bhilwara Group & Bakshi Group of Companies for 5 yrs. and has been teaching for last 11 years as Faculty in reputed Business Schools. She also worked as Project Officer with NITRA and ATIRA at Ahmedabad for 5 years.
She has published several research papers in National & International journals

**Piyush Kumar GUPTA, Verification and Emulation Methodology and Tools Group Manager, ST Microelectronics Pvt Ltd Noida.**
**Email:** piyush-kumar.gupta@st.com

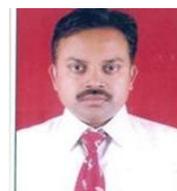

Piyush Kumar GUPTA has B.Tech. Degree in Electronics and Communication from reputed institute, Ram Manohar Lohia University Faizabad. He has around 15 years of work experience, and currently leading Central Functional Verification and Emulation Methodology Team to define and deploy new flow/tools in Verification and Emulation. Work closely with global ST sites to effectively collaborate on various activities which involved many key research areas e.g. Unified UVM Architecture for simulation and Emulation, UPF support on Emulation, Automated Assertions generation, Graph based Test Generation etc.
His expertise is in Dynamic/Formal verification and Emulation R&D.

**Sachish Dhar DWIVEDI, SDS Group, STMicroelectronics Pvt. Ltd.**
**Email:** sachish.dwivedi@st.com

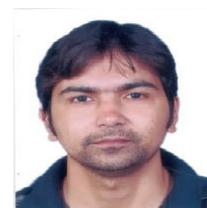

Sachish Dhar DWIVEDI has more than 8 years of direct experience in Industry. He has done Masters in Technology from Motilal Nehru National Institute of Technology, Allahabad in 2004. He is mainly responsible for SOC/IP Verification/Pre-Silicon Validation using Advanced Hardware Emulation Techniques. His main areas of interest are to develop Emulation methodologies, Evaluation of new tools/techniques and support to various emulation based activities.